\documentclass[review]{elsarticle}

\usepackage{hyperref}

\journal{Journal of \LaTeX\ Templates}

\usepackage{graphicx}

\usepackage{amssymb}
\usepackage{latexsym}

\begin{document}

\begin{frontmatter}

\title{
Imaginary-field-driven phase transition
for
the $2$D
Ising antiferromagnet:
A fidelity-susceptibility approach
}

\author{Yoshihiro Nishiyama}
\address{Department of Physics, Faculty of Science,
Okayama University, Okayama 700-8530, Japan}




\begin{abstract}

The square-lattice Ising antiferromagnet 
subjected to the imaginary magnetic field
$H=i \theta T /2 $
with the ``topological'' angle $\theta$ and temperature $T$
was investigated by means of  the transfer-matrix method.
Here,
as a probe to detect the order-disorder phase transition,
we adopt an extended version of the fidelity susceptibility
$\chi_F^{(\theta)}$, which makes sense even for such a non-hermitian transfer matrix.
As a preliminary survey,
for an intermediate value of $\theta$, 
we examined the finite-size-scaling behavior of $\chi_F^{(\theta)}$,
and found a pronounced signature for the criticality;
note that the magnetic susceptibility exhibits a weak (logarithmic) singularity
at the N\'eel temperature.
Thereby, we turn to the analysis of the power-law singularity of the phase boundary 
at  $\theta=\pi$.
With $\theta-\pi$ scaled properly,
the $\chi_F^{(\theta)}$ data are cast into the crossover scaling
formula, indicating that the phase boundary is shaped concavely.
Such a feature makes a marked contrast to that of the mean-field theory.

\end{abstract}


\end{frontmatter}


\section{\label{section1}Introduction}

The fidelity $F$ is given by the overlap,
$F=| \langle \theta | \theta + \Delta \theta \rangle |$,
between the ground states
with the proximate interaction parameters,
$\theta$ and $\theta +\Delta \theta$;
here, the symbol $| \theta \rangle$ denotes the ground-state vector
for a certain  parameter $\theta$.
The concept of fidelity has been developed in the course of the
studies on the quantum dynamics
\cite{Uhlmann76,Jozsa94,Peres84,Gorin06}.
Meanwhile,
it turned out that the fidelity is sensitive to the quantum phase transition
\cite{Quan06,Zanardi06,Zanardi07,You07,HQZhou08,You11,Rossini18}.
Actually, 
the fidelity susceptibility
$ - \frac{1}{N} \partial_{\Delta \theta}^2  F|_{\Delta \theta = 0}$
with the number of lattice points $N$
exhibits a pronounced signature for the criticality
\cite{Zanardi07,You07,Albuquerque10},
as compared to those of the conventional 
quantifiers such as the specific heat and magnetic susceptibility.
Moreover, it does not rely on any presumptions as to the order parameter
involved. 
Clearly, the fidelity fits the exact-diagonalization scheme,
with which an explicit expression for $| \theta \rangle$ is readily available.
It has to be mentioned, however, that the fidelity is
accessible via the quantum Monte Carlo method
\cite{Albuquerque10,Schwandt09,Grandi11,Wang15}
as well as the experimental observations
\cite{Zhang08,Kolodrubetz13,Gu14}.

Then, there arose a problem
whether the concept of fidelity is applicable to the transfer-matrix simulation scheme
\cite{Zhou08,Sirker10}.
In Ref. \cite{Sirker10},
the concept of fidelity was extended so as to treat the 
quantum transfer matrix for 
the $XXZ$ spin chain at finite temperatures.
The quantum transfer matrix takes a non-symmetric form,
although the matrix elements are {\em real}.
Hence, the concept of fidelity does not apply to the transfer-matrix
simulation scheme straightforwardly.
To circumvent the difficulty,
there was proposed the following extended formula for the fidelity 
\cite{Schwandt09,Sirker10}
\begin{equation}
\label{extended_fidelity}
F= \sqrt{ 
\frac{
 [	{\bf v}_L(\theta+\Delta \theta) \cdot {\bf v}_R(\theta) ][
	{\bf v}_L(\theta) \cdot {\bf v}_R(\theta+\Delta \theta)]
}{
[	{\bf v}_L(\theta) \cdot {\bf v}_R(\theta)    ][
{\bf v}_L(\theta+\Delta \theta) \cdot {\bf v}_R(\theta+\Delta \theta)]
}
} 
 .
\end{equation}
Here, the symbols
${\bf v}_{L,R}(\theta)$
denote the left ($L$) and right ($R$) eigenvectors
satisfying
\begin{eqnarray}
^{t} {\bf v}_L  (\theta) T(\theta) &=& \lambda_1 {}^t {\bf v}_L(\theta)    \\
T(\theta){\bf v}_R(\theta) &=& \lambda_1 {\bf v}_R(\theta)   
,
\end{eqnarray}
with the largest eigenvalue $\lambda_1$
for the transfer matrix $T(\theta)$
\cite{Forcrand18}, and
the variable $\theta$ stands for a certain system parameter.
Provided that $T(\theta)$ is hermitian (like the Hamiltonian),
the above expression (\ref{extended_fidelity})
recovers the above-mentioned formula $|\langle \theta| \theta+\Delta \theta\rangle |$
because of
$^t {\bf v}_L(\theta) \propto \langle \theta|$
and
$ {\bf v}_R(\theta) \propto | \theta \rangle$.
According to the elaborated simulation study
\cite{Sirker10},
the
extended fidelity works 
as in the case of the Hamiltonian formalism,
and the fidelity captures
a notable signature of the criticality
for the finite-temperature quantum $XXZ$ spin chain.

In this paper,
we adopt the aforementioned expression 
(\ref{extended_fidelity})
for $F$
in order
to
treat
the case of the {\em non-hermitian} transfer matrix.
For that purpose,
we consider
the square-lattice Ising antiferromagnet
subjected to the imaginary magnetic field.
We show that the expression 
(\ref{extended_fidelity}) works
in the non-hermitian-transfer-matrix case.  
As a preliminary survey,
we investigate the order-disorder phase transition
via the fidelity susceptibility.
Thereby, we 
examine how the critical branch ends up,
as the imaginary magnetic field is strengthened.

The Hamiltonian
${\cal H}$
for the square-lattice Ising antiferromagnet
under the imaginary magnetic field
is given by the expression
\begin{equation}
\label{Hamiltonian}
{\cal H}=
J \sum_{\langle ij \rangle} \sigma_i \sigma_j -H\sum_{i} \sigma_i 
.
\end{equation}
Here, the Ising spin $\sigma_i =\pm 1$ is placed at each square-lattice point $i$.
The summation $\sum_{\langle ij\rangle}$
runs over all possible nearest-neighbor pairs
$\langle ij \rangle$.
The magnetic field $H$ is set to a pure imaginary value,
$H = i \theta T /2$,
with the 
``topological'' angle $\theta$ and temperature $T$,
and likewise,
the reduced coupling constant $K=J/T$
with the antiferromagnetic interaction $J$ 
is introduced.
In the presence of the magnetic field,
the antiferromagnetic model (\ref{Hamiltonian})
exhibits quite different characters 
\cite{Kim04}
from the
ferromagnetic counterpart.
The latter has been investigated extensively
\cite{Lee52,Garcia-Saez92}
in the context of the Lee-Yang zeros 
of the partition function.
Nevertheless, the imaginary magnetic field,
the so-called topological
$\theta$-term, renders a severe sign problem
\cite{Azcoiti11,Azcoiti13},
for which
the Monte Carlo method does not work very efficiently.
In this paper, we surmount the difficulty by means of the transfer-matrix method
\cite{Forcrand18}
with the aid of the above-mentioned fidelity susceptibility.
Rather intriguingly, the imaginary magnetic field $\theta=\pi$ has 
a physical interpretation in the $K<0$ side.
According to the duality argument \cite{Suzuki90,Lin88} 
at
$\theta=\pi$, the system is mapped to the (fully) frustrated magnet
in the $K<0$ side.

\begin{figure}
\includegraphics[width=120mm]{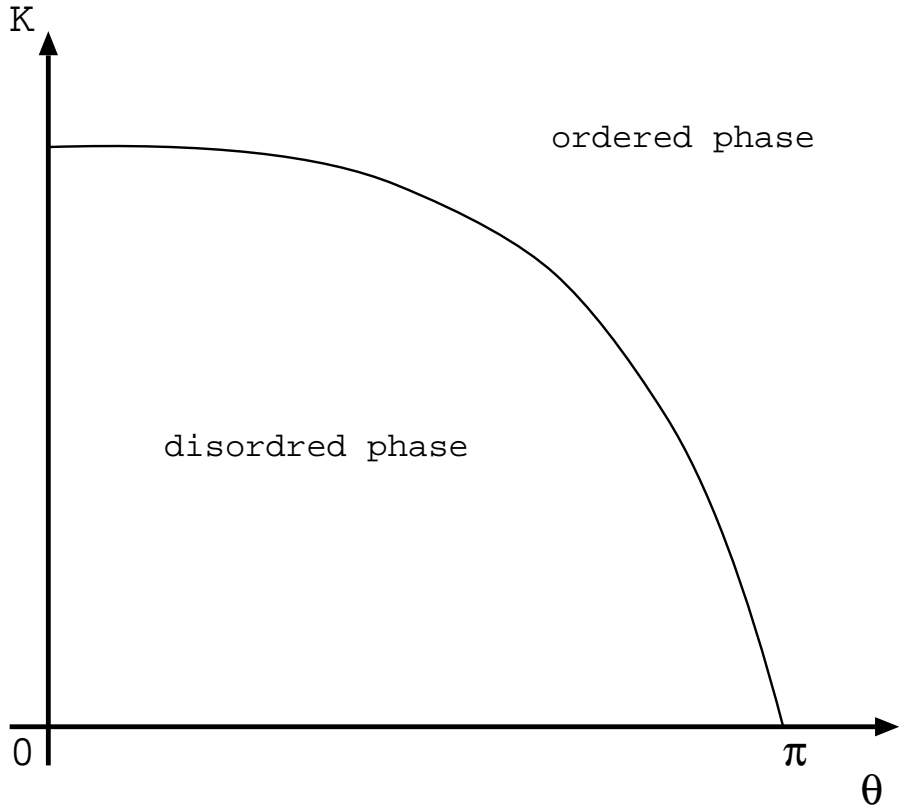}%
\caption{\label{figure1}
A schematic drawing of the phase diagram for the square-lattice 
Ising antiferromagnet under the imaginary magnetic field
$H=i \theta T/2$
(\ref{Hamiltonian})
is presented 
\cite{Matreev08,Azcoiti17,Azcoiti11,McCoy67}.
The symbols, $K(=J/T)$ and $\theta$, denote the
reduced coupling constant and the 
``topological'' angle, respectively.
In contrast to the ferromagnet counterpart
\cite{Lee52},
the order-disorder phase boundary extends into the $\theta>0$ regime,
and eventually, it terminates at $\theta=\pi$.
The end-point singularity is one of our concerns.
}
\end{figure}

\begin{figure}
\includegraphics[width=120mm]{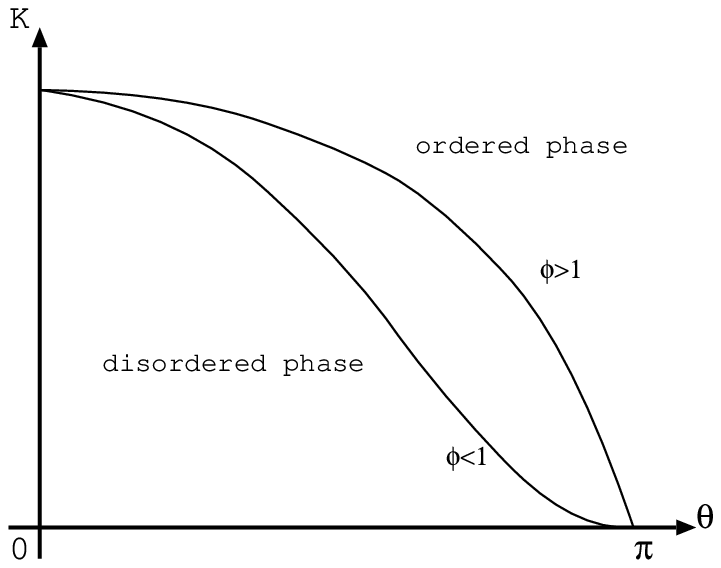}%
\caption{\label{figure2}
The power-law singularity of
the phase boundary toward the end-point $\theta=\pi$
is described by the crossover exponent $\phi$.
The mean-field theory predicts 
\cite{Azcoiti11}
a convex curvature, $\phi(=1/2)<1$.
On the contrary, the series of data points 
with the partition-function-zeros \cite{Matreev08}
and cumulant-expansion \cite{Azcoiti17}
methods seem to suggest 
a concave curvature, $\phi>1$.
}
\end{figure}

A 
schematic $\theta$-$K$ phase diagram 
for the system (\ref{Hamiltonian}) is presented
in Fig. \ref{figure1}.
The overall characteristic  was elucidated by the
partition-function-zeros \cite{Matreev08}
and cumulant-expansion \cite{Azcoiti17}  studies.
Additionally,
along $\theta=\pi$, 
rigorous information
\cite{Lee52,McCoy67}
as well as the analytic-continuation Monte Carlo
results  
\cite{Azcoiti11}
are available.
These results
suggest
that
the order-disorder-phase boundary 
extends into an intermediate-$\theta (\ne 0)$ regime
(unlike the ferromagnetic counterpart),
and it terminates
at the multicritical point $(\theta,K)=(\pi,0)$
eventually.
The character of the multicriticality is one of our concerns.
Actually, the mean-field analysis \cite{Azcoiti11}
indicates that the phase boundary is of a convex function
in the vicinity of
the multicritical point $\theta=\pi$;
see Fig. \ref{figure2}.
Such a feature
suggests that the crossover exponent takes a small value, $\phi(=1/2)<1$;
On the one hand, the series of the data points 
of the simulation studies \cite{Matreev08,Azcoiti17}
display a concave curvature, $\phi>1$.
In this paper, the end-point singularity is explored
quantitatively by casting the fidelity-susceptibility data
into the crossover scaling formula
with $\theta-\pi$ scaled carefully.

The rest of this paper is organized as follows.
In Sec. \ref{section2},
following the explanation of the simulation scheme,
the numerical results are presented.
Related preceding studies are also recollected.
In Sec. \ref{section3},
we address the summary and discussions.

\section{\label{section2}Numerical results}

In this section,
we present the numerical results
for the square-lattice Ising antiferromagnet 
under the imaginary magnetic field (\ref{Hamiltonian}).
We employed the
transfer-matrix method 
\cite{Forcrand18}.
The transfer-matrix elements are 
given by the row-to-row statistical weight
along the transfer-matrix-strip direction.
The transfer-matrix strip width extends up to $L \le 20$,
and
the periodic-boundary condition was imposed.
Thereby,
the fidelity susceptibility was calculated by
the formula
\begin{equation}
\label{fidelity_susceptibility}
\chi_F^{(\theta)} = - \frac{1}{L} \partial_{\Delta\theta}^2 F(\theta,\theta + \Delta \theta)|_{\Delta \theta =0}
,
\end{equation}
with the extended 
\cite{Sirker10,Schwandt09}
fidelity $F$ defined by Eq. (\ref{extended_fidelity}).
We show that the fidelity susceptibility exhibits
a pronounced signature
for the order-disorder phase transition;
actually,
the ordinary uniform magnetic susceptibility exhibits
a weak (logarithmic) singularity 
at the N\'eel temperature
\cite{Fisher60,Kaufman87}.
Rather confusingly,
the denominator $L$ of the above expression (\ref{fidelity_susceptibility})
differs from
that of
the ordinary quantum-mechanical
treatment, $N$.
It
comes from the transfer-matrix-slice size $L$,
which has to be regarded as the normalization factor in this case.

\subsection{\label{section2_1}
Finite-size-scaling
analysis
of the fidelity susceptibility $\chi_F^{(\theta)}$ at $\theta=\pi/2$}

In this section, we investigate the critical behavior of 
the fidelity susceptibility $\chi_F^{(\theta)}$ with $\theta$ fixed to
an intermediate value, $\pi/2$.
As mentioned in Introduction,
the order-disorder phase boundary for the antiferromagnet extends into the finite-$\theta$
regime in contrast to the ferromagnetic counterpart.

\begin{figure}
\includegraphics[width=120mm]{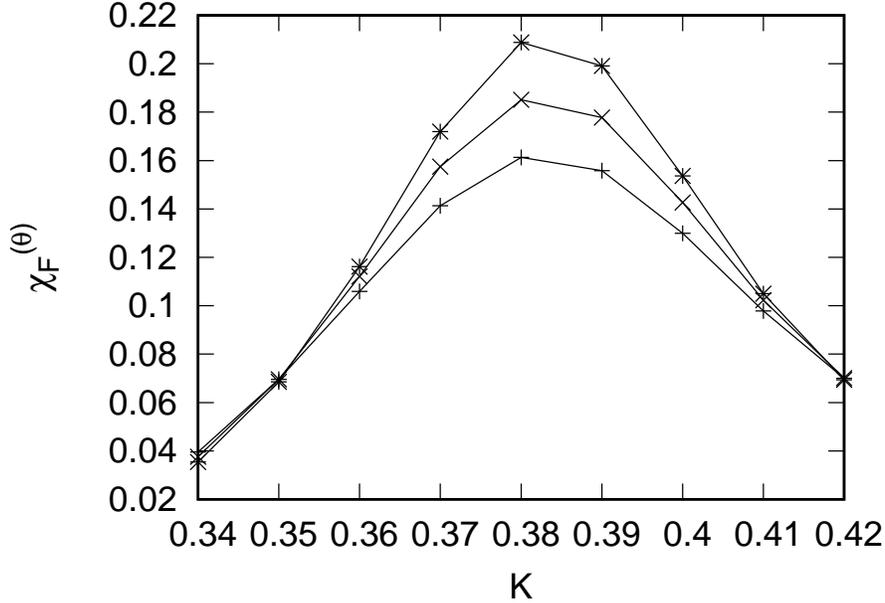}%
\caption{\label{figure3}
The fidelity susceptibility
$\chi_F^{(\theta)}$
(\ref{fidelity_susceptibility})
is plotted for the
reduced coupling constant $K$
with the fixed $\theta=\pi/2$ and
various 
system sizes,
($+$) $L=16$,
($\times$) $18$, and
($*$) $20$.
The fidelity susceptibility exhibits a notable peak around
$K \approx 0.38$;
note that the ordinary {\em uniform} susceptibility for the antiferromagnet
exhibits a weak (logarithmic) singularity.
}
\end{figure}

To begin with, in Fig. \ref{figure3},
we present the fidelity susceptibility $\chi_F^{(\theta)}$ (\ref{fidelity_susceptibility})
for the reduced coupling constant $K$
and various system sizes,
($+$) $L=16$,
($\times$) $18$, and
($*$) $20$.
Here, the imaginary magnetic field is fixed to an intermediate value $\theta=\pi/2$.
The fidelity susceptibility exhibits a notable peak around $K \approx 0.38$,
which indicates an onset of the order-disorder phase transition.

\begin{figure}
\includegraphics[width=120mm]{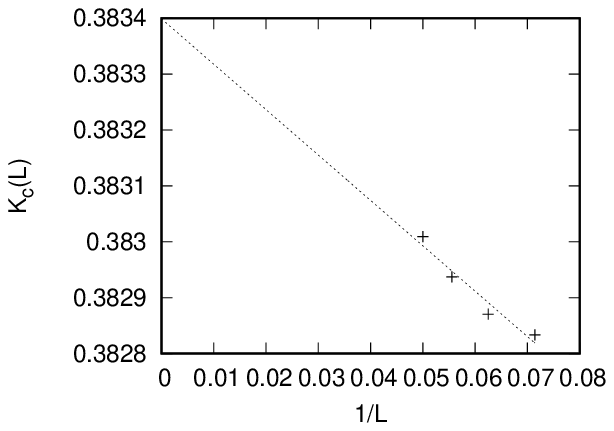}%
\caption{\label{figure4}
The approximate critical point 
$K_c(L)$ 
(\ref{approximate_critical_point})
is plotted for $1/L$ with the fixed $\theta=\pi/2$,
and $L=14,16,\dots,20$.
The least-squares fit yields an estimate 
$K_c=0.38340(9)$ 
in the thermodynamic limit $L\to\infty$.
A possible systematic error is considered in the text.
Related preceding results are recollected in Table \ref{table1}.
}
\end{figure}

In Fig. \ref{figure4},
we present the approximate critical point $K_c(L)$ for $1/L$
with the fixed $\theta=\pi/2$ and various system sizes, $L=14,16,\dots,20$.
Here,
the approximate critical point
$K_c(L)$
denotes the maximal point for the fidelity susceptibility;
\begin{equation}
\label{approximate_critical_point}
\partial_K \chi_F^{(\theta)}|_{K=K_c(L)}=0
.
\end{equation}
The least-squares fit to the data in Fig. \ref{figure4}
yields an estimate 
$K_c=0.38340(9)$ in the thermodynamic 
limit $L\to\infty$.
Alternatively,
we arrived at $K_c=0.38315(6)$,
replacing the abscissa scale
with $1/L^2$ in the least-squares-fit analysis.
The deviation between them, $\approx 4 \cdot 10^{-4}$,
appears to dominate the least-squares-fit error, $ \approx 9 \cdot 10^{-5}$.
Hence,
considering the former as an indicator for a possible systematic error,
we estimate the critical point as
\begin{equation}
\label{critical_point}
K_c=0.3834(4).
\end{equation}

\begin{table}
\caption{
Related studies of the square-lattice Ising antiferromagnet
under the imaginary magnetic field $H=i\theta T/2$
(\ref{Hamiltonian}) are recollected.
So far, a variety of techniques,
such as
finite-cluster's
exact partition function 
\cite{Matreev08},
first-$k$-term cumulant expansion
\cite{Azcoiti17}, 
and transfer-matrix diagonalization
(this work),
have been utilized.
In order to detect the order-disorder phase transition,
there have been proposed a number of quantifiers,
such as
partition-function zeros, staggered-magnetization-fluctuation 
singularity
($\partial_\theta \langle m_s^2\rangle$),
and fidelity susceptibility, respectively.
As a reference, the transition point 
$K_c|_{\theta=\pi/2}$ at an intermediate $\theta=\pi/2$,
is shown for each study;
the cumulant-expansion results are read off from Fig. 11 of Ref. \cite{Azcoiti17}.
}
\label{table1}       
\begin{tabular}{lll}
\hline\noalign{\smallskip}
 method & quantifier & transition point $K_c|_{\theta=\pi/2}$ \\
\noalign{\smallskip}\hline\noalign{\smallskip}
partition function \cite{Matreev08} & partition-function zeros & $0.382(\approx \frac{\ln 4.6}{4})$ \\  
first-$k$-term cumulant expansion \cite{Azcoiti17} & $\partial_\theta \langle m_s^2\rangle$ & $0.359$ ($k=8$), $0.341$ ($k=4$) \\
transfer matrix (this work) & fidelity susceptibility &  $0.3834(4)$  \\
\noalign{\smallskip}\hline
\end{tabular}
\end{table}

So far, the critical point $K_c|_{\theta=\pi/2}$
at $\theta=\pi/2$
has been estimated with the partition-function-zeros 
\cite{Matreev08}
and first-$k$-term-cumulant-expansion \cite{Azcoiti17} methods.
We recollected them in
Table \ref{table1};
here, the cumulant-expansion estimates $K_c|_{\theta=\pi/2}$ 
are read off
from Fig. 11 of Ref. \cite{Azcoiti17}.
In the respective studies,
as a quantifier for the phase transition, 
the accumulation of the
partition-function zeros and singularity of the 
staggered-magnetization fluctuations,
$\partial_\theta \langle m_s^2\rangle $, were utilized. 
The former approach yields an estimate
$K_c|_{\theta=\pi/2} =0.382(\approx \frac{\ln 4.6}{4})$, whereas the latter reported
$K_c|_{\theta=\pi/2}=0.359$ and $0.341$ for $k=8$ and $4$, respectively.
Our result $K_c|_{\theta=\pi/2}=0.3834(4)$ 
[Eq. (\ref{critical_point})] supports these elaborated pioneering studies.
A benefit of the $\chi_F^{(\theta)}$-mediated approach \cite{Yu09}
is that, as shown in Fig. \ref{figure4},
the finite-size data $K_c(L)$, albeit with rather restricted $L$,
converge rapidly to the thermodynamic limit.

\begin{figure}
\includegraphics[width=120mm]{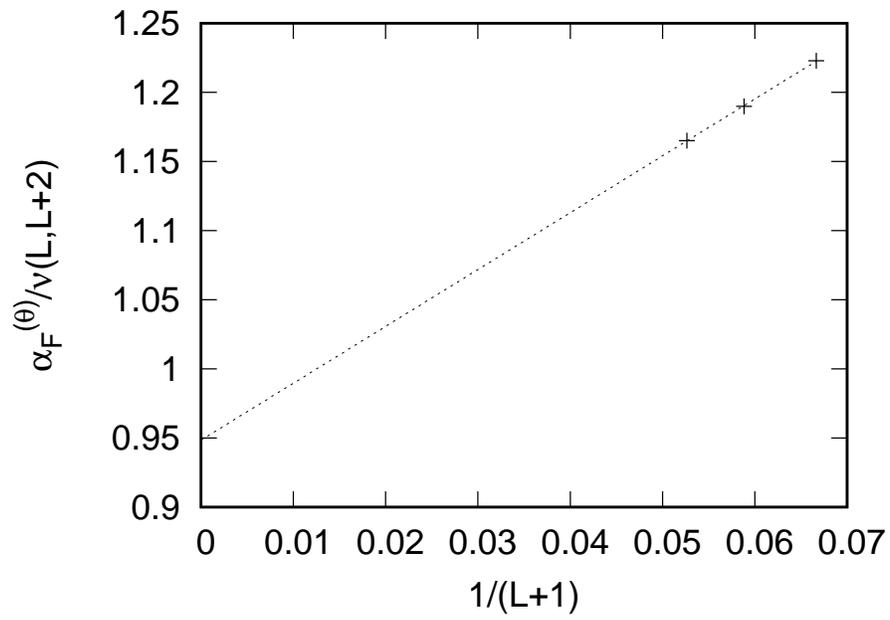}%
\caption{\label{figure5}
The approximate critical exponent 
$\frac{\alpha_F^{(\theta)}}{\nu}(L,L+2)$
(\ref{approximate_critical_exponent})
is plotted for $1/(L+1)$ with the fixed $\theta=\pi/2$,
and $L=14,16,18$.
The least-squares fit yields an estimate 
$\alpha/\nu=0.949(4)$
in the thermodynamic limit
$L\to\infty$.
A possible systematic error is considered in the text.
}
\end{figure}

We turn to the analysis of the critical exponent
$x=\alpha_F^{(\theta)}/\nu$,
namely, the scaling dimension
for the fidelity susceptibility \cite{Albuquerque10}.
Here, the exponent $\alpha_F^{(\theta)}$
describes the singularity of
the fidelity susceptibility,
$\chi_F^{(\theta)} \sim |K-K_c|^{-\alpha_F^{(\theta)}}$,  
whereas 
the index $\nu$ denotes the correlation-length critical exponent,
$\xi \sim |K-K_c|^{-\nu}$.
In Fig. \ref{figure5},
we present the approximate critical exponent
$\frac{ \alpha_F^{(\theta)} }{ \nu}(L,L+2)$
for $1/(L+1)$ with $\theta$ fixed to an intermediate value $\pi/2$,
and $L=14,16,18$.
Here, the approximate critical exponent is given by
the formula
\begin{equation}
\label{approximate_critical_exponent}
\frac{ \alpha_F^{(\theta)}  }{  \nu } (L,L')=
\frac{
\ln \chi_F^{(\theta)}(L)|_{K=K_c(L)} -\ln \chi_F^{(\theta)}(L')|_{K=K_c(L')}
}{  \ln L - \ln L'  }
     ,
\end{equation}
for a pair of system sizes, $(L,L')$.
The least-squares fit to the data in Fig. \ref{figure5}
yields an estimate 
$\alpha/\nu=0.949(4)$ in the thermodynamics limit 
$L\to\infty$.
Alternatively,
we arrived at an estimate
$\alpha/\nu=1.070(3)$,
replacing the abscissa scale with $1/L^2$
in the extrapolation scheme.
Considering the deviation between them,
$\approx 0.12$, as a possible systematic error,
we estimate the critical exponent as
\begin{equation}
\label{critical_exponent}
x=\alpha_F^{(\theta)}    /\nu=0.95(12)
.
\end{equation}

According to
the scaling relation \cite{Albuquerque10}
\begin{equation}
\label{scaling_relation}
\alpha_F^{(\theta)}/ \nu =
\gamma_{af}/\nu +1
,
\end{equation}
with the magnetic-susceptibility critical exponent for the antiferromagnet $\gamma_{af}$,
our result $\alpha_F^{(\theta)}/\nu=0.95(12)$ 
[Eq. (\ref{critical_exponent})] yields
an estimate
\begin{equation}
\label{critical_exponent_2}
\gamma_{af}/\nu=-0.05(12)
.
\end{equation}
The result is accordant with that of
the two-dimensional Ising antiferromagnet
$\gamma_{af}=0$ (logarithmic)
\cite{Fisher60,Kaufman87}.

A few remarks are in order.
First,
it has been reported that
the recent series-expansion ``data
are not extensive enough to calculate the critical exponents''
\cite{Azcoiti17}.
The present $\chi_F^{(\theta)}$-mediated approach 
captures an evidence
that the criticality belongs to the two-dimensional
Ising universality class.
This point is further pursued in Sec \ref{section2_2}.
Last, a key ingredient is that $\chi_F^{(\theta)}$'s singularity is stronger 
than
that of the ordinary quantifiers such as the specific heat and magnetic susceptibility;
note that both quantifiers exhibit weak (logarithmic) divergences
at the transition point for the two-dimensional Ising antiferromagnet.
As shown in Sec. \ref{section2_3},
the fidelity susceptibility exhibits an even stronger singularity 
right
at the multicritical point
$(\theta,K)=(\pi,0)$.

\subsection{\label{section2_2}
Scaling plot of the fidelity susceptibility $\chi_F^{(\theta)}$
at $\theta=\pi/2$}

In this section,
we display the scaling plot for $\chi_F^{(\theta)}$,
based on the scaling formula
\cite{Albuquerque10}
\begin{equation}
\label{scaling_formula}
\chi_F^{(\theta)}=L^{x} f \left((K-K_c)L^{1/\nu}\right)
    ,
\end{equation}
with $\chi_F^{(\theta)}$'s scaling dimension $x=\alpha_F^{(\theta)}/\nu$, and a non-universal 
scaling function $f$.
The scaling parameters, $K_c$ 
(\ref{critical_point})
and $\alpha_F^{(\theta)}/\nu$
(\ref{critical_exponent}),
are fed into the formula (\ref{scaling_formula})
in order to crosscheck the analyses in Sec. \ref{section2_1}.
The index $\nu$ remains adjustable to be fixed in the subsequent survey.

\begin{figure}
\includegraphics[width=120mm]{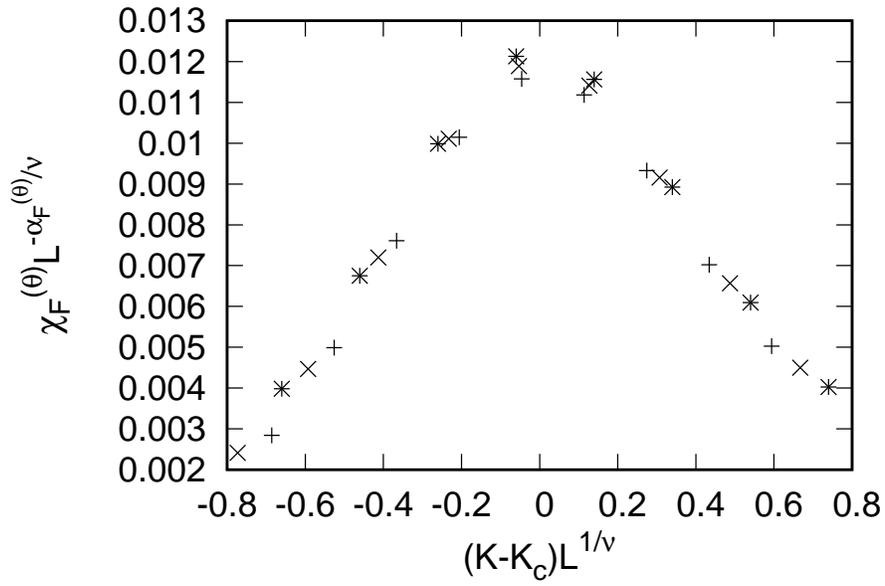}%
\caption{\label{figure6}
The scaling plot, $(K-K_c)L^{1/\nu}$-$\chi_F^{(\theta)}L^{-\alpha_F^{(\theta)}/\nu}$,
is presented with the fixed $\theta=\pi/2$
for 
various system sizes,
($+$) $L=16$,
($\times$) $18$, and
($*$) $20$.
Here, the scaling parameters are set to 
$K_c=0.3834$ [Eq. (\ref{critical_point})],
$\alpha_F^{(\theta)}/\nu=0.95$ [Eq. (\ref{critical_exponent})],
and $\nu=1$ (2D-Ising universality class);
the present scaling analysis is based on the formula (\ref{scaling_formula}).
}
\end{figure}

In Fig. \ref{figure6}, we present the scaling plot,
$(K-K_c)L^{1/\nu}$-$\chi_F^{(\theta)}L^{-\alpha_F^{(\theta)}/\nu}$,
with the fixed $\theta=\pi/2$
for various system sizes, 
($+$) $N=16$
($\times$) $18$,
and 
($*$) $20$.
Here, we assumed the two-dimensional Ising universality class,
$\nu=1$,
and the other
scaling parameters
are set to
$K_c=0.3834$ [Eq. (\ref{critical_point})]
and
$\alpha_F^{(\theta)}/\nu=0.95$ 
[Eq. (\ref{critical_exponent})].

The data in Fig. \ref{figure6}
collapse into a scaling function $f$ satisfactorily.
The result
validates the scaling analyses in Sec. \ref{section2_1}
and the proposition $\nu=1$ (2D Ising universality).
Therefore, recollecting
the aforementioned estimate
 $\gamma_{af}/\nu=-0.05(12)$
[Eq. (\ref{critical_exponent_2})],
we 
confirm that the criticality belongs to the two-dimensional
Ising universality class.

Last, we address a remark.
As shown in Fig. \ref{figure6},
the fidelity susceptibility exhibits a notable peak around the critical point.
Namely, 
the fidelity susceptibility picks up the singular part
out of non-singular (background) contributions.
Such an elimination of non-singular part is 
significant so as to
make a reliable scaling analysis of the criticality.
Actually,
the fidelity-susceptibility-mediated approach
\cite{Yu09} 
succeeded in the analysis of
the 2D quantum criticality 
via the exact diagonalization method with rather  restricted system sizes.
Such a benefit seems to be retained
for the non-hermitian-transfer-matrix formalism.

\subsection{\label{section2_3}Crossover scaling plot of the
fidelity susceptibility $\chi_F^{(\theta)}$
around $\theta=\pi$}

In this section, 
based on the crossover scaling theory
\cite{Riedel69,Pfeuty74},
we
investigate the end-point singularity of the phase boundary
toward $\theta = \pi$.
For that purpose,
we introduce yet another parameter,
namely,
the distance from the multicritical point,
$\delta \theta =\pi - \theta$,
accompanied with the
crossover critical exponent $\phi$.
Thereby, 
the scaling formula
takes an extended expression
\begin{equation}
\label{extended_scaling_formula}
\chi_F^{(\theta)}
=
L^{\dot{x}} g
\left(
\left(K-K_c(\theta)\right)L^{1/\dot{\nu}},\delta \theta L^{\phi/\dot{\nu}}
\right)
,
\end{equation}
with 
the fidelity-susceptibility and correlation-length critical exponents,
$\dot{\alpha}_F^{(\theta)}$
and
$\dot{\nu}$, respectively, right at the multicritical point $\delta \theta= 0$,
and
a non-universal scaling function $g$.
As in Eq. (\ref{scaling_formula}),
the index $\dot{x}$ denotes
the scaling dimension for the fidelity susceptibility 
 $\dot{x}=\dot{\alpha}_F^{(\theta)} / \dot{\nu}$
at $\delta \theta=0$.

Before commencing the scaling analyses,
the values of the critical indices,
$\dot{\nu}$ and
$\dot{x}=\dot{\alpha}_F^{(\theta)}/\dot{\nu}$, are fixed.
The multicriticality occurs at the
high temperature limit $K_c=0$,
where the correlation length does not develop,
and 
the finite-size behavior obeys
the putative scaling law
$\dot{\nu}=1$ \cite{Fisher82,Challa86,Campostrini15}.
As in Eq. (\ref{scaling_relation}),
the index $\dot{x}$ satisfies
\cite{Albuquerque10}
the scaling relation
$\dot{x}=\dot{\alpha}_F/\dot{\nu}
=\dot{\gamma}/\dot{\nu} +1+1$,
where the third term $1$ comes from the coefficient of the susceptibility formula
$T^{-1}\partial_\theta^2 \ln Z$
($Z$: partition function).
This relation admits
$\dot{x}=4.5$ 
because of 
the susceptibility exponent
$\dot{\gamma}=2.5$ at $\theta=\pi$ \cite{Matreev95},
and the aforementioned $\dot{\nu}=1$.
The index
$\phi$ remains adjustable 
so as to be determined in the subsequent analyses. 
It is to be noted that
the crossover exponent $\phi$
is relevant to the power-law singularity
of the phase boundary
\cite{Riedel69,Pfeuty74},
$K_c(\theta) \sim |\pi-\theta|^{1/\phi}$;
see
Fig. \ref{figure2} as well.
As mentioned in Introduction,
the mean-field theory admits a convex curvature $\phi(=1/2) < 1$
around $\theta=\pi$.

\begin{figure}
\includegraphics[width=120mm]{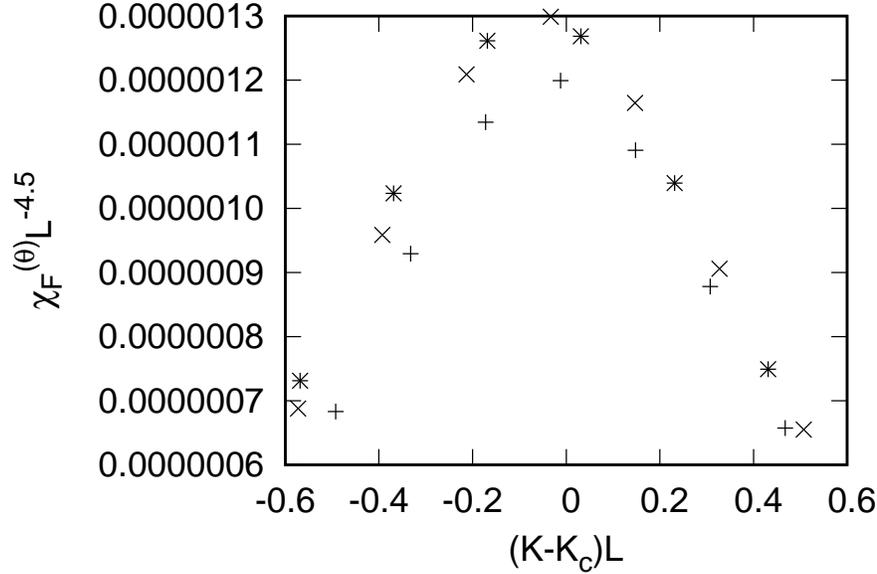}%
\caption{\label{figure7}
The crossover scaling plot,
$(K-K_c(\theta))L$-$\chi_F^{(\theta)}L^{-4.5}$,
is presented with the fixed $\delta \theta L^\phi=314$ 
(the second argument of Eq. (\ref{extended_scaling_formula}))
for
various system sizes,
($+$) $L=16$,
($\times$) $18$, and
($*$) $20$.
Here, the crossover exponent is set to
an optimal value
$\phi=2$.
}
\end{figure}

In Fig. \ref{figure7},
we present the scaling plot, 
$(K-K_c(\theta))L$-$\chi_F^{(\theta)}L^{-4.5}$,
for the various system sizes,
($+$) $L=16$,
($\times$) $18$, and
($*$) $20$.
Here, the second argument of the scaling function $g$ 
in
Eq. (\ref{extended_scaling_formula}) 
is fixed to 
$\delta \theta L^{\phi}=314$ under the proposition,
$\phi=2$,
and the critical point $K_c(\theta)$ was determined via the same scheme as in Sec.
\ref{section2_1}.
The crossover-scaled data in Fig. \ref{figure7}
collapse into the scaling function $g$
satisfactorily.
This result indicates that the choice 
$\phi=2$ is a plausible one.

\begin{figure}
\includegraphics[width=120mm]{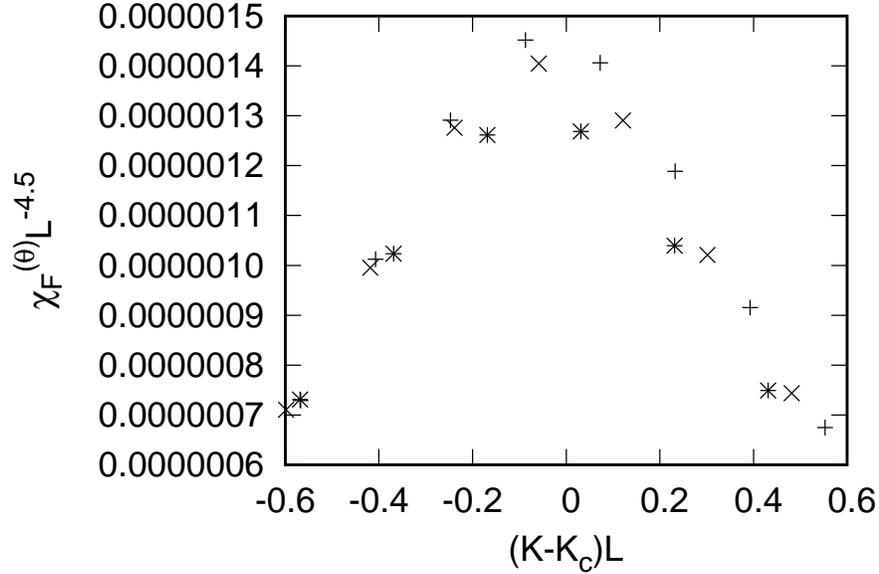}%
\caption{\label{figure8}
The crossover scaling plot,
$(K-K_c(\theta))L$-$\chi_F^{(\theta)}L^{-4.5}$,
is presented with the fixed $\delta \theta L^\phi=94.8$ 
(the second argument of Eq. (\ref{extended_scaling_formula}))
for
various system sizes,
($+$) $L=16$,
($\times$) $18$, and
($*$) $20$.
Here, the crossover exponent is set to
$\phi=1.6$.
For small $\phi$,
the right-side slope gets resolved,
as compared to that of Fig. \ref{figure7};
see the separation between $L=18$ ($\times$)
and $20$ ($*$), in particular.
}
\end{figure}

\begin{figure}
\includegraphics[width=120mm]{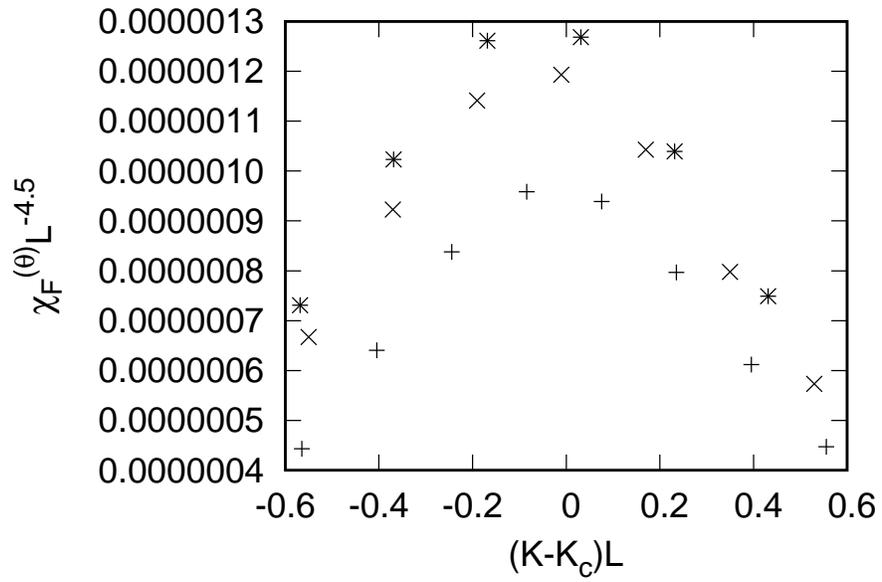}%
\caption{\label{figure9}
The crossover scaling plot,
$(K-K_c(\theta))L$-$\chi_F^{(\theta)}L^{-4.5}$,
is presented with the fixed $\delta \theta L^\phi=1041$ 
(the second argument of Eq. (\ref{extended_scaling_formula}))
for
various system sizes,
($+$) $L=16$,
($\times$) $18$, and
($*$) $20$.
Here, the crossover exponent is set to
$\phi=2.4$.
For large $\phi$,
the left-side slope gets scattered,
as compared to that of Fig. \ref{figure7}.
}
\end{figure}

As a reference, we made the similar scaling analyses
for various values of $\phi$.
In Fig. \ref{figure8}, we present the scaling plot, 
$(K-K_c(\theta))L$-$\chi_F^{(\theta)}L^{-4.5}$,
with the fixed $\delta \theta L^\phi=94.8$
under the setting, $\phi=1.6$;
the symbols are the same as those of Fig. \ref{figure7}.
The scaled data become scattered, as compared to those of Fig.
\ref{figure7}; particularly,
the data constituting the  right-side slope and hilltop
get resolved.
The mean-field case $\phi=1/2(<2)$ belongs this category,
and the data should become even scattered.
Likewise,
in Fig. \ref{figure9}, we display the scaling plot,
$(K-K_c(\theta))L$-$\chi_F^{(\theta)}L^{-4.5}$, 
with $\delta \theta L^\phi=1041$ under the proposition,
$\phi=2.4$;
the symbols are the same as those of Fig. \ref{figure7}.
For such large $\phi$, on the contrary,
the left-side-slope data become scattered.
As a result,
we conclude that
the crossover exponent lies within
\begin{equation}
\label{crossover_critical_exponent}
\phi =2.0(4)
.
\end{equation}

A few remarks are in order.
First,
the underlying mechanism behind the crossover scaling plot, Fig. \ref{figure7},
differs from that of the 
fixed-$\theta$
scaling, Fig. \ref{figure6}.
Actually, the former scaling dimension $\dot{x}=4.5$ is much lager
that the latter $x=0.95(12)$, Eq. (\ref{critical_exponent}),
and hence, the data collapse in Fig. \ref{figure7} is by no means accidental.
Second,
our result $\phi[=2.0(4)] >1$, Eq. (\ref{crossover_critical_exponent}),
suggests that the end-point state at $(\theta,K)=(\pi,0) $
is sensitive to the external-field
($\delta \theta$-driven) perturbation rather than the
thermal ($K$-driven) one.
In other words, the magnetic fluctuation is enhanced toward the end-point.
Such a feature is consistent with the duality theory
\cite{Suzuki90,Lin88},
which states that the system with  $K \le 0$ ($\theta=\pi$)
reduces to the {\em fully}-frustrated model.
Therefore,
the index $\phi > 1$ reflects a precursor to entering into the frustrated magnetism.
Because the mapping \cite{Suzuki90,Lin88} is validated only in two dimensions,
it is reasonable that the mean-field result $\phi=1/2$ 
\cite{Azcoiti11}
does not 
capture this character.
Last, 
even for
such an exotic phase transition, the fidelity-susceptibility approach works.
The fidelity susceptibility does not rely on any
{\it ad hoc} presumptions as to the order parameter involved.   

\section{\label{section3}
Summary and discussions}

The square-lattice Ising antiferromagnet 
subjected to the imaginary magnetic field $H=i \theta T/2$
(\ref{Hamiltonian})
was investigated with the transfer-matrix method.
As a probe to detect the phase transition,
we utilized the extended version
\cite{Schwandt09,Sirker10}
of the fidelity
(\ref{extended_fidelity}),
which is applicable to the non-hermitian-transfer-matrix formalism.
As a demonstration, we calculated the fidelity susceptibility
$\chi_F^{(\theta)}$ 
(\ref{fidelity_susceptibility})
for an intermediate value
of $\theta=\pi/2$, and analyzed the order-disorder phase transition.
The transition point 
$K_c=0.3834(4)$ [Eq. (\ref{critical_point})] appears to support the preceding analyses
\cite{Matreev08,Azcoiti17}.
Moreover, we found
that the critical indices,
$\gamma_{af}/\nu=-0.05(12)$ [Eq. (\ref{critical_exponent_2})] and 
$\nu=1$ (Sec. \ref{section2_2}),
agree with those of
the two-dimensional Ising universality class.
Note that so far, it has been reported \cite{Azcoiti17} that the ``data
are not extensive enough to calculate the critical exponents''
as to the critical branch.
Here, a key ingredient is that
$\chi_F^{(\theta)}$'s scaling dimension, $x=\alpha_F^{(\theta)}/\nu=0.95(12)$
[Eq. (\ref{critical_exponent})], is larger than that of the magnetic susceptibility,
$\gamma_{af}/\nu=-0.05(12)$ (logarithmic \cite{Fisher60,Kaufman87}),
and 
$\chi_F^{(\theta)}$-aided analysis picks up \cite{Yu09}
the singularity out of the background contributions clearly.
We then turn to the analysis of
the end-point singularity of the phase boundary at $\theta=\pi$.
With $\pi-\theta$ scaled properly,
the $\chi_F^{(\theta)}$ data are cast into the crossover scaling theory 
(\ref{extended_scaling_formula}).
We attained a data collapse through adjusting the crossover exponent to
$\phi=2.0(4)$ 
[Eq. (\ref{crossover_critical_exponent})].
This result $\phi>1$ indicates that
the phase boundary is formed concavely
around the end-point $\theta=\pi$
in marked contrast to the
mean-field \cite{Azcoiti11} prediction, $\phi=1/2$.
In other words, the multicritical point
$(\theta,K)=(\pi,0)$ is sensitive to the external-field perturbation
rather than the thermal one.

As a matter of fact, 
at $\theta=\pi$ ($K \le 0$),
the model (\ref{Hamiltonian}) reduces to the fully-frustrated model
\cite{Suzuki90,Lin88} through the duality transformation.
Hence, it is reasonable that the magnetism at $\theta=\pi$
($K \le 0$) is sensitive to the external-field perturbation.
In this sense, the end-point singularity is regarded as a precursor to
the frustrated magnetism.
Because the duality theory is validated only in two dimensions,
the mean-field theory does not capture this character.
It would be tempting to apply the present scheme to the $K<0$ side so as to
elucidate the $\theta$-induced frustrated magnetism \cite{Suzuki90} via 
the probe $\chi_F^{(\theta)}$.
This problem is left for the future study.

\section*{References}


\end{document}